\documentclass[a4paper,11pt]{article}
\usepackage{jcappub}
\usepackage{float} 
\usepackage{xspace}

\newcommand\edit[1]{\textcolor{black}{#1}}
\newcommand\editt[1]{\textcolor{black}{#1}}
\newcommand{\CNB}{C$\nu$B\xspace}

\title{Multi-Messenger Astrophysics with the Cosmic Neutrino Background}
\author[1]{Christopher G.\ Tully}
\author[2]{and Gemma Zhang}
\affiliation[1]{Department of Physics, Princeton University, Princeton, New Jersey 08544, USA}
\affiliation[2]{Department of Physics, Harvard University, Cambridge, Massachusetts 02138, USA}
\emailAdd{cgtully@princeton.edu}
\emailAdd{yzhang7@g.harvard.edu}

\abstract{
The massive neutrinos of the Cosmic Neutrino Background (\CNB) are fundamental ingredients of the radiation-dominated early universe and are important non-relativistic probes of the large-scale structure formation in the late universe.  The dominant source of anisotropies in the neutrino flux distribution on the sky are highly amplified integrals of metric perturbations encountered during the non-relativistic phase of the \CNB.  This paper numerically compares the line-of-sight methods for computing \CNB anisotropies with the Einstein-Boltzmann hierarchy solutions in linear theory for a range of neutrino masses.  Angular power spectra are computed that are relevant to a future polarized tritium target run of the PTOLEMY experiment.  Correlations between the \CNB sky maps and galactic survey data are derived using line-of-sight techniques and discussed in the context of multi-messenger astrophysics.
}
\begin{document}

\maketitle
\flushbottom

\section{Introduction}
Two of the most important predictions of the standard cosmological model are the Cosmic Neutrino Background (\CNB) and the Cosmic Microwave Background (CMB). The former was produced from neutrino decoupling approximately one second after the Big Bang while the latter was produced from photon decoupling approximately 380,000 years after the Big Bang. In the present day, the monopole temperature of the \CNB is predicted to be around 1.95~K, which is $\left(\frac{4}{11}\right)^{\frac{1}{3}}$ of the CMB temperature due to the electron-positron annihilation that occurred between neutrino decoupling and photon decoupling. Over the last several decades, the anisotropies in the CMB have been precisely measured and extensively analyzed. On the other hand, although the \CNB can potentially shed light on the universe in a much earlier era than the CMB, because of the challenges in experimentally measuring relic neutrinos, the importance of studying the \CNB anisotropies has often been overlooked. The PTOLEMY experiment is the first of its kind aimed at detecting the \CNB~\cite{ptolemy2019}. The prospect of a direct detection of relic neutrinos necessitates more thorough theoretical studies of the \CNB than currently exist in literature. 

The \CNB and CMB were produced as radiation in the early universe, and one may naturally assume that they share many characteristics. However, this assumption turns out to be reasonable only for massless neutrinos. Massive neutrinos that initially behave as radiation and later turn non-relativistic generate significant differences in the \CNB compared to the CMB.  As an example, the distance to the last scattering surface of the \CNB has been shown to be much closer to us than that of the CMB, on the scale of a few Gpc, depending on the neutrino mass~\cite{dodelson2009}.  In general, the non-relativistic transition of massive neutrinos introduces growth in the non-uniformity of the \CNB similar to the large-scale structure formation of Cold Dark Matter (CDM), but on length scales that are unique to the \CNB temperature and neutrino masses.

This paper compares the \CNB anisotropies calculated from the Boltzmann hierarchy equations with those from the line-of-sight method and reconciles the differences in the results presented by Refs.~\cite{hannestad2010} and \cite{michney2007} in Section~\ref{sec:calculations}. We show consistent results from the two methods through numertical computations in Section~\ref{sec:results}.  The finite mass of neutrinos results in an amplified anisotropic angular power spectra in the low multipole moments: for a neutrino mass of $0.01$~eV, for example, the anisotropies are amplified by more than 100~times compared to the massless case. We discuss the experimental implications of our results in Section~\ref{sec:exp} and conclude in Section~\ref{sec:conclusion}.

\section{\CNB Anisotropy Calculations} 
\label{sec:calculations}
The \CNB anisotropies are calculated from the perturbation to the neutrino Fermi-Dirac distribution. For massive neutrinos that are relativistic at neutrino decoupling, the perturbed Fermi-Dirac distribution is given as: 
\begin{align}
    f(x^{i}, q, n_j, \tau) = f_0(q) (1 + \Psi(x^{i}, q, n_j, \tau)) = \frac{1}{e^{\frac{q}{T_0}}+1}(1 + \Psi(x^{i}, q, n_j, \tau)), 
    \label{eq:FD}
\end{align}
where we use notations following Ref.~\cite{ma1995}. Note in particular that $q = ap$ is the comoving momentum and $T_0 = a T$ is the present day \CNB temperature with $a(\tau)$ being the scale factor in the conformal Newtonian gauge
\begin{align}
    ds^2 = a(\tau)^2(-(1+2\psi)d\tau^2 + (1-2\phi)dx^idx_i)
    \label{eq:newtongauge}
\end{align}
where $\phi$ and $\psi$ are the metric perturbations.
The unperturbed distribution $f_0(q) = (e^{\frac{\epsilon}{T_0}}+1)^{-1}$ where $\epsilon = \sqrt{q^2 + (am)^2}$ is approximated as $f_0(q) = (e^{\frac{q}{T_0}}+1)^{-1}$ at neutrino decoupling when $a \simeq 10^{-10}$ and $q \gg am$, and this distribution in phase space is frozen once neutrinos decouple. The perturbation $\Psi$ to the Fermi-Dirac distribution for neutrinos can be solved from the collisionless Boltzmann equation which in the conformal Newtonian gauge is written as: 
\begin{align}
    \frac{\partial \Psi}{\partial \tau} + i \frac{q}{\epsilon} (\vec{k} \cdot \hat{n}) \Psi + \frac{d\ln{f_0}}{d\ln{q}}(\dot{\phi} - i \edit{\frac{\epsilon}{q}} (\vec{k} \cdot \hat{n}) \psi) = 0
    \label{eq:boltzmann1}
\end{align}
where $\vec{k}$ is the wavenumber of the Fourier mode of the perturbation and the line-of-sight direction is $\hat{n}$.
To obtain the temperature power spectra, we calculate the temperature perturbations by writing the perturbed distribution as $f(q) = (e^{\frac{q}{T_0(1+\Delta)}}+1)^{-1}$ which allows us to expand to linear order and write the temperature perturbation in terms of $\Psi$ as
\begin{align}
    \Delta = -\left(\frac{d\ln{f_0}}{d\ln{q}}\right)^{-1} \Psi .
    \label{eq:delta_def}
\end{align}
Here, we outline two ways to solve for $\Psi$ from the Boltzmann equation: (1)~solving the Boltzmann hierarchy as detailed in Ref.~\cite{hannestad2010}; (2)~calculating a line-of-sight integral of the Boltzmann equation and applying appropriate approximations as detailed in Ref.~\cite{michney2007}. The \CNB anisotropy power spectra calculated in the two references for massive neutrinos appear inconsistent by several orders of magnitude depending on the neutrino mass with Ref.~\cite{hannestad2010} predicting the massive neutrino anisotropies in a range between the CMB and the CDM large-scale structure while Ref.~\cite{michney2007} predicts massive neutrinos are more uniform than the CMB. We will apply several corrections to the line-of-sight integral method presented in the reference and show that the two methods in fact yield consistent results for massive neutrinos lighter than 0.1~eV with the line-of-sight approximations beginning to become inaccurate for heavier masses.

The Boltzmann hierarchy method writes the solution as an expansion in the Legendre polynomials: $\Psi = \sum_{l=0}^{\infty} (-i)^l (2l+1) P_l(\hat{k} \cdot \hat{n}) \Psi_l$. Substituting this expansion into the Boltzmann equation gives us an infinite set of equations that can be solved numerically. Ref.~\cite{oldengott2015} discusses the Boltzmann hierarchy for neutrinos in more detail. The solutions to $\Psi_l$ in turn gives the solutions to the temperature perturbations $\Delta_l(q)$ for comoving momentum $q$ of a given neutrino species. The perturbations solved by the Boltzmann hierarchy method can be obtained from cosmology codes such as \texttt{CLASS} and \texttt{CAMB}~\cite{class2011,2011ascl.soft02026L}. To express our results in the form of power spectra, we use the convention that $\langle \delta_{\mathbf{k}} \delta_{\mathbf{k}'} \rangle = (2\pi)^3 P_0(k)$ where $\delta_{\mathbf{k}}$ is Fourier transformed primordial curvature perturbation, and the primordial power spectrum is given as 
\begin{align}
    P_0 = 2\pi^2 A_s k^{-3} \left(\frac{k}{k_{\mathrm{pivot}}}\right)^{n_s-1} \exp{\left(\frac{\alpha_s}{2}\ln{\left(\frac{k}{k_{\mathrm{pivot}}}\right)^2}\right)}
\end{align}
where $n_s$ is the scalar spectral index, and we assume the tilt running to be $\alpha_s = 0$ and a Harrison-Zel'dovich-Peebles spectrum where $n_s = 1$ in the following discussion. The value of $k_{\mathrm{pivot}}$ is chosen arbitrarily and is typically chosen as $0.05~\mathrm{Mpc}^{-1}$, but for our scale invariant spectrum, the value of $k_{\mathrm{pivot}}$ is inconsequential. The angular power spectrum is therefore: 
\begin{align}
    C_l(q) = (4 \pi)^2 T_0^2 \int \frac{k^2 dk}{(2\pi)^3} P_0(k) \Delta_l^2(q) = 4\pi A_s T_0^2 \int d\ln{k} \Delta_l^2(q) \ .
    \label{eq:clq_def}
\end{align}
Note that this is the angular power spectrum for a given comoving neutrino momentum $q$. We will discuss the $q$-independent angular power spectrum in a later part of this paper.

In the line-of-sight integral method, we start by assuming that $\psi = \phi$ and rewriting the Boltzmann equation in terms of $\Delta$ and $\Gamma \equiv \frac{d\ln{f_0}}{d\ln{q}}$: 
\begin{align}
    \partial_{\lambda} (\Gamma \Delta) + \Gamma \left(\frac{\epsilon^2}{q^2} \partial_{\lambda} + \left(1 + \frac{\epsilon^2}{q^2}\right)\partial_{\tau}\right)\phi = 0, 
    \label{eq:boltzmann2}
\end{align}
where we take $-\partial_{\lambda} = \partial_{\tau} + \frac{q}{\epsilon} \hat{n} \cdot \vec{\nabla}$. \edit{Note that there is a caveat to the $\psi = \phi$ assumption for massless neutrinos, which we will address near the end of our calculations.} Then we can write Eq.~\eqref{eq:boltzmann2} again as 
\begin{align}
    \partial_{\lambda} (\Gamma \Delta) + \partial_{\lambda} \left(\frac{\epsilon^2}{q^2} \Gamma\phi\right) - \partial_{\lambda} \left(\frac{\epsilon^2}{q^2} \Gamma\right)\phi + \Gamma\left(1 + \frac{\epsilon^2}{q^2}\right)\partial_{\tau}\phi = 0. 
    \label{eq:boltzmann3}
\end{align}
Then performing the line-of-sight integral from neutrino decoupling to present day of Eq.~\eqref{eq:boltzmann3} gives us the present day perturbation:
\begin{align}
    \Delta_0 = -\frac{\epsilon^2}{q^2} \phi_0 + \frac{\Gamma_{\mathrm{dec}}}{\Gamma_0} \left(\Delta_{\mathrm{dec}} + \frac{\epsilon^2}{q^2} \Bigg |_{\mathrm{dec}}\phi_{\mathrm{dec}}\right) + \frac{1}{\Gamma_0} \int_{\mathrm{dec}}^{0} \left(\partial_{\lambda} \left(\frac{\epsilon^2}{q^2} \Gamma\right)\phi -\Gamma\left(1 + \frac{\epsilon^2}{q^2}\right)\partial_{\tau}\phi\right) d\lambda. 
    \label{eq:perturbation}
\end{align}
In Eq.~\eqref{eq:perturbation}, the first term only contributes to the monopole and hence can be neglected in our anisotropy calculation and we take $\frac{\epsilon}{q} \sim 1$ at decoupling for relativistic neutrinos. Following the discussion in Ref.~\cite{michney2007}, we also take $\Delta_{\mathrm{dec}} + \phi_{\mathrm{dec}} = \frac{1}{2}\phi_{\mathrm{dec}}$. Therefore, Eq.~\eqref{eq:perturbation} simplifies to 
\begin{align}
    \Delta_0 = \frac{1}{2}\frac{\Gamma_{\mathrm{dec}}}{\Gamma_0} \phi_{\mathrm{dec}} + \frac{1}{\Gamma_0} \int_{\mathrm{dec}}^{0} \left(\partial_{\lambda} \left(\frac{\epsilon^2}{q^2} \Gamma\right)\phi - \Gamma\left(1 + \frac{\epsilon^2}{q^2}\right)\partial_{\tau}\phi\right) d\lambda. 
\end{align}
We note that $\Gamma$ is a constant of time and therefore the factors of $\Gamma$ cancel out in the expression for $\Delta_0$: 
\begin{align}
    \Delta_0 = \frac{1}{2} \phi_{\mathrm{dec}} + \int_{\mathrm{dec}}^{0} \left(\partial_{\lambda} \left(\frac{am}{q}\right)^2 \phi - \left(2 + \left(\frac{am}{q}\right)^2\right)\partial_{\tau}\phi\right) d\lambda.
    \label{eq:delta0}
\end{align}
We can further simplify Eq.~\eqref{eq:delta0} by evaluating the partial derivative
\begin{align}
    \partial_{\lambda}
    \left(\frac{am}{q} \right)^2 
    \edit{= -\partial_{\tau}
    \left(\frac{am}{q} \right)^2 = -}2 \left(\frac{am}{q} \right)^2 a H
    \label{eq:partiallamba}
\end{align}
with
\begin{align}
    H = H_0 \sqrt{\Omega_m/a^3 + \Omega_r/a^4 + \Omega_\Lambda}
    \label{eq:H}
\end{align}
for the present day Hubble constant, $H_0$, and fractional energy densities relative to critical density of matter ($\Omega_m$), radiation ($\Omega_r$) and dark energy, assuming $\Omega_\Lambda = 1 - \Omega_m - \Omega_r$.
Therefore, using the definition of $\Delta_l$ from Ref.~\cite{schoneberg2018} and the definition of $C_l(q)$ given in Eq.~\eqref{eq:clq_def}, we find the $q$-dependent angular power spectrum to be 
\begin{align}
    C_l(q) & = 4\pi T_0^2 A_s \int  d\ln{k} \nonumber \\
    & \left(\frac{1}{2} \phi_{\mathrm{dec}} j_l(k \chi(z_{\mathrm{dec}}))
    + \int_{\mathrm{dec}}^{0} \left(
    \edit{-} 2 \left(\frac{am}{q} \right)^2 \left( a H \right)
    \phi -\left(2 + \left(\frac{am}{q}\right)^2\right)\partial_{\tau}\phi\right) j_l(k \chi(z_{\lambda})) d\lambda\right)^2 \\
    &= \edit{4\pi T_0^2 A_s \int  d\ln{k} \nonumber} \\
    & \edit{\left(\frac{1}{2} \phi_{\mathrm{dec}} j_l(k \chi(z_{\mathrm{dec}}))
    + \int_{\tau_\mathrm{dec}}^{\tau_0} \left(
    2 \left(\frac{am}{q} \right)^2 \left( a H \right)
    \phi + \left(2 + \left(\frac{am}{q}\right)^2\right)\partial_{\tau}\phi\right) j_l(k \chi(z_{\tau})) d\tau\right)^2}.
    \label{eq:clq}
\end{align}
Here, we note that $j_l$ are Bessel functions of the first kind and the normalization $A_s$ is related to the normalization $A$ in Ref.~\cite{michney2007} as $A_s T_0^2 = 4\pi A$. We use $\chi$ to denote the comoving distance traveled by massive neutrinos from a time specified by redshift $z$ to the present and is given by Ref.~\cite{dodelson2009} as
\begin{align}
    \chi(z) = \int_{t(z)}^{t_0} \frac{c q}{a(t) \epsilon} dt = \int_{a(z)}^{1} \frac{c}{a^2 H(a)}\frac{q}{\epsilon} da,
    \label{eq:chi_def}
\end{align}
where $c$ is the speed of light and $t_0$ is the age of the universe today. Note that $\chi(z_{\mathrm{dec}})$ is then the comoving distance to the neutrino last scattering surface. Compared to the results in Ref.~\cite{michney2007}, our $\Delta_0$ expression has two additional terms containing a factor of $\left(\frac{am}{q}\right)^2 \sim \left(\frac{am}{ap}\right)^2 \sim \frac{1}{v^2}$. At late times, we have $am \gg q$, so these terms dominate, and as massive neutrinos slow down and become non-relativistic, the anisotropies become amplified by this $\frac{1}{v^2}$ factor. This factor is briefly discussed in Ref.~\cite{hannestad2010} and resolves one of the main disagreements between Refs.~\cite{michney2007} and \cite{hannestad2010}. The amplification of \CNB angular power spectrum at low multipoles is shown in Fig.~\ref{fig:compare}.

\edit{To obtain Eq.~\eqref{eq:clq}, we assumed $\psi = \phi$ in the Boltzmann equation to simplify our calculations. This assumption is reasonable for massive neutrinos, but in the case of massless neutrinos, we find that this assumption has a non-negligible effect on the angular power spectra. Therefore, relaxing the $\psi = \phi$ assumption and then taking $m = 0$, we get the $q$-dependent angular power spectrum of massless neutrinos as:
\begin{align}
    C_l(q) &= 4\pi T_0^2 A_s \int  d\ln{k} \left(\left(\psi_{\mathrm{dec}} - \frac{1}{2}\phi_{\mathrm{dec}}\right)j_l(k \chi(z_{\mathrm{dec}}))
    + \int_{\tau_\mathrm{dec}}^{\tau_0} (\partial_{\tau} \psi + \partial_{\tau} \phi) j_l(k\chi(z_{\tau})) d\tau\right)^2.
    \label{eq:clq_massless}
\end{align}
Note that without the $\psi = \phi$ assumption, the first term in the $\int d\ln k$ integrand becomes $\psi_{\mathrm{dec}} - \frac{1}{2}\phi_{\mathrm{dec}}$ instead of $\frac{1}{2}\phi_{\mathrm{dec}}$. This term dominates the massless neutrino power spectra, so we need to account for the non-vanishing $\phi-\psi$ difference.}

We summarize the above calculations by highlighting the main modifications made to the line-of-sight calculations in Ref.~\cite{michney2007}: (1)~we take the unperturbed Fermi-Dirac distribution as $(e^{\frac{q}{T}}+1)^{-1}$ instead of $(e^{\frac{\epsilon}{T}}+1)^{-1}$ for neutrinos that are relativistic at decoupling and assume that such distribution remains constant after neutrino decoupling, so our definition of the temperature perturbation $\Delta$ contains terms that get amplified at late times; (2)~the distance to last scattering surface argument in Eq.~\eqref{eq:clq} is the distance traveled by massive neutrinos instead of the distance traveled by light. 
\edit{In Section~\ref{sec:results}, we will show that our numerical results from the line-of-sight calculations
agree well with the Boltzmann hierarchy results. 
\editt{The results in Ref.~\cite{2015Alipour} disagree with Refs.~\cite{hannestad2010} and \cite{michney2007}, but this discrepancy appears to be largely due to the incorrect application of a visibility function in the massive case.}
The Boltzmann hierarchy calculations are independent of the assumptions made in our line-of-sight calculations, so this agreement strengthens the validity of the various assumptions we have discussed above. }

Direct detection rates of relic neutrinos through neutrino capture on nuclei, as discussed in Section~\ref{sec:exp}, are not strongly sensitive to the incoming non-relativistic momentum of the captured neutrinos.  This behavior follows from the product of the capture cross section times velocity being constant for neutrino kinetic energies below approximately 1~keV~\cite{Cocco:2007za}.  Therefore, starting from a cross-spectrum of the angular power for comoving neutrino momenta $q_1$ and $q_2$ for a given neutrino species, we have
\begin{align}
    C_l(q_1,q_2) = 4\pi A_s T_0^2 \int d\ln{k} \Delta_l(k,q_1)\Delta_l(k,q_2).
    \label{eq:clq1q2_def}
\end{align}
We can remove the $q$-dependence from $\Delta_l(k,q)$ by integrating over the normalized differential neutrino number density $n(q) dq/n_\nu$ to give
\begin{align}
    \bar{\Delta}_l(k) & = \nonumber \\
    & \edit{\int \left(\frac{1}{2} \phi_{\mathrm{dec}} j_l(k \chi(z_{\mathrm{dec}}))
    + \int_{\tau_{\mathrm{dec}}}^{\tau_0} \left(
    2 \left(\frac{am}{q} \right)^2 \left( a H \right)
    \phi + \left(2 + \left(\frac{am}{q}\right)^2\right)\partial_{\tau}\phi\right) j_l(k \chi(z_{\tau})) d\tau\right)} \nonumber \\
    & \times \left( \frac{2}{3 \zeta(3) T_0^3} \right)
    \frac{q^2 d q}{e^{\frac{q}{T_0}} + 1},
    \label{eq:deltal}
\end{align}
where $\zeta$ is the Riemann zeta function and comes from the normalization of the neutrino number density, where
\begin{align}
    n_\nu = \int_0^\infty n(q) dq = \int_0^\infty \frac{2}{(2\pi)^3} (4 \pi) \frac{q^2 dq}{e^{\frac{q}{T_0}} + 1} = \frac{3 \zeta(3)}{2 \pi^2}T_0^3
    \label{eq:nnu}
\end{align}
for one species of neutrinos isotropically distributed.
The $q$-independent angular power spectrum for neutrino capture experiments is therefore
\begin{align}
    C_l = 4\pi A_s T_0^2 \int d\ln{k} \bar{\Delta}_l(k)^2.
    \label{eq:cl}
\end{align}
A similar $q$-bin averaging is proposed in Ref.~\cite{hannestad2010}, but assumes that measurements are performed in $q$-bins and then averaged.  In Eq.~\eqref{eq:cl}, the $q$-dependence of the observable is integrated out separately for two different points on the sky.

\section{Analysis \& Results}
\label{sec:results}
We compare the results of these two methods of calculating the \CNB anisotropies numerically by plotting their angular power spectra. We obtain the temperature and metric perturbations from running the \texttt{CLASS} functions in CO$N$CEPT~\cite{class2011, dakin2019}. The \texttt{CLASS} software uses the Boltzmann hierarchy method to solve for the temperature perturbations which are then used to calculate our angular power spectrum. We use the metric perturbations at an array of redshifts to calculate the line-of-sight integrals. \edit{The results for massless and massive neutrinos are computed using Eq.~\eqref{eq:clq_massless} and Eq.~\eqref{eq:clq}, respectively.} The numerical arrays for the comparisons that follow are provided by \texttt{CLASS} with the computational details available online.\footnote{  \url{https://github.com/gemyxzhang/cnb-anisotropies.git}}

\begin{figure}[H]
    \centering
    \includegraphics[width=1\linewidth]{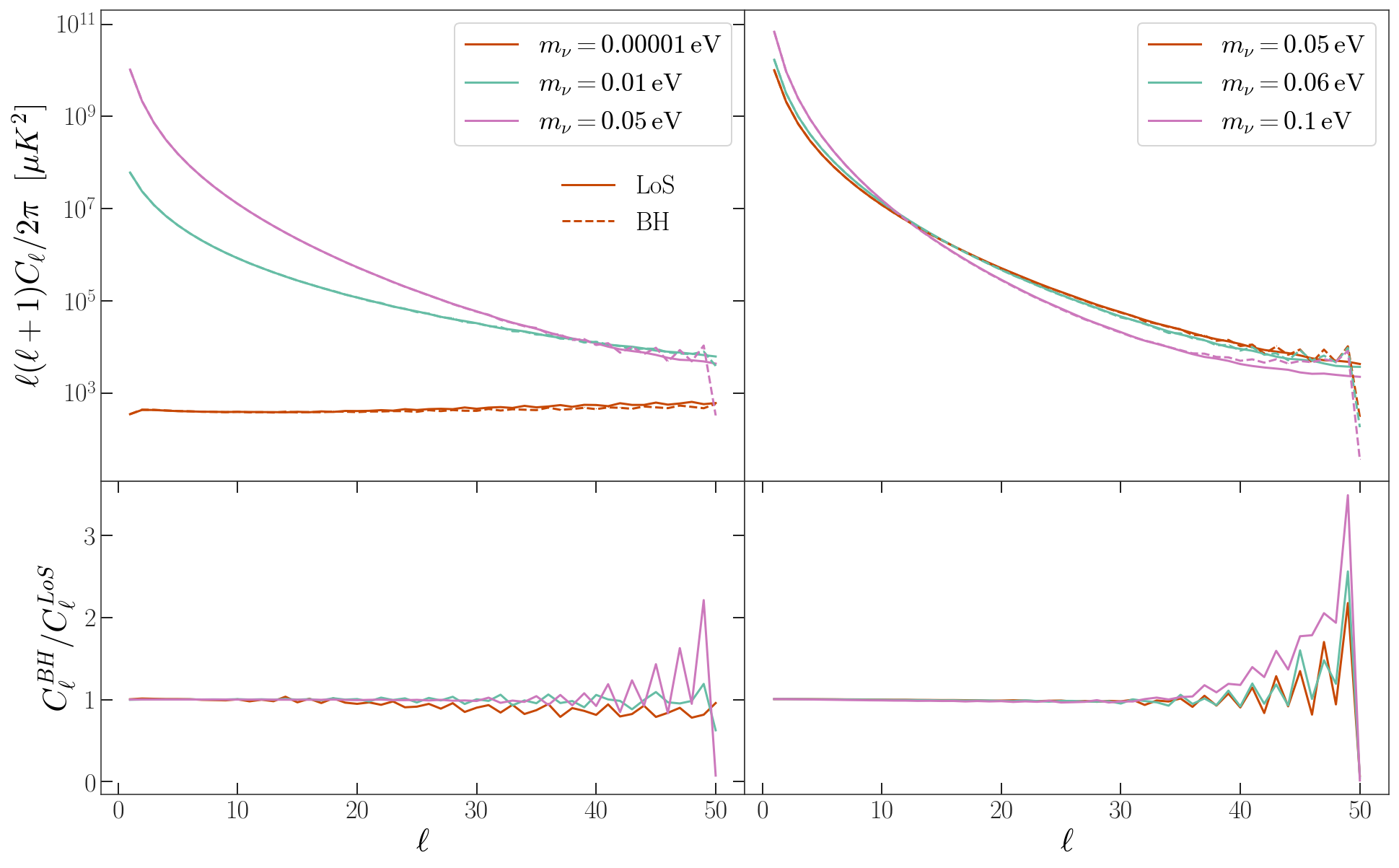}
    \caption{\textbf{Upper panel}: The angular power spectra calculated using the Boltzmann hierarchy method (dotted) compared with those calculated using the line-of-sight integral method (solid) for neutrino masses ({\it left}) of 0.00001~eV (red), 0.01~eV (green), and 0.05~eV (purple) and with the absolute masses shifted up by 0.05~eV ({\it right}). The integral over $k$ is performed over a $k$ range of $(10^{-4}, 10^{-1})$~Mpc$^{-1}$ in logarithmic bins. The $C_l$ shown in the figure are the $C_l(q)$ for $q = 3 T_0$. \textbf{Lower panel}: Ratios of the Boltzmann hierarchy method and line-of-sight integral method angular power spectra in the upper panel.}
    \label{fig:compare}
\end{figure}

We compare in Fig.~\ref{fig:compare} the power spectra produced from the Boltzmann hierarchy (BH) method and the line-of-sight (LoS) integral method for an $l$ range of $[1,50]$, two sets of neutrino masses (\{0.00001~eV, 0.01~eV, 0.05~eV\} and \{0.05~eV, 0.06~eV, 0.1~eV\}), and for a $k$ range of $(10^{-4}, 10^{-1})$~Mpc$^{-1}$. The two methods agree to within 10\% over several orders of magnitude.  Deviations begin to become visible for $l>30$ and masses of $0.1$~eV where the upper-limit of the $k$ integral significantly impacts the computed $C_l$.  Most notably, the $m=0.1$~eV curve in Fig.~\ref{fig:compare} crosses the $m=0.05$~eV and $m=0.06$~eV curves because of the finite $k$ integral, where the $m=0.1$~eV neutrino perturbations arise mainly from $k$ larger than $10^{-1}$~Mpc$^{-1}$. \edit{The LoS and BH results of the almost massless ($0.00001$~eV) neutrinos appear to differ more than those of the massive neutrinos. This is because the massless neutrino angular power spectra, devoid of the non-relativistic terms, are more sensitive than their massive counterparts to the numerical approximations we made due to software limitations. Nevertheless, we note that our results for the almost massless neutrinos agree with previous results in Refs.~\cite{hu1995, michney2007}. }

\begin{figure}[H]
    \centering
    \includegraphics[width=1\linewidth]{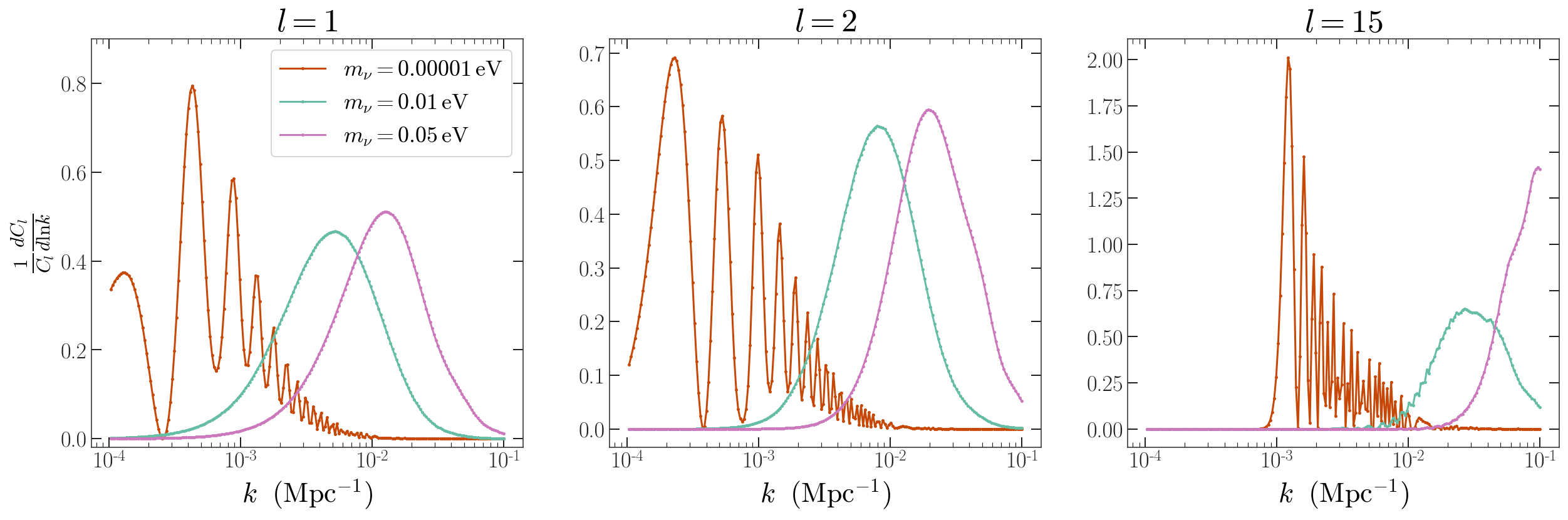}
    \caption{The fractional contributions to $C_l$ for $q=3 T_0$ as a function of $k$-mode for the dipole $l=1$ ({\it left}), quadrupole $l=2$ ({\it middle}) and high $l$-mode $l=15$ ({\it right}) for neutrino masses of 0.00001~eV (red), 0.01~eV (green), and 0.05~eV (purple).  The oscillations for the nearly massless case mark the proper time distance to the Hubble crossing.  The massive neutrino $k$-mode sensitivities are peaked in different ranges depending the $l$-mode and mass.}
    \label{fig:dCldk}
\end{figure}

The differential contribution to $C_l$ ($l=1,2,15$) as a function of $k$ mode is shown in Fig.~\ref{fig:dCldk} for several neutrino masses.  From Fig.~\ref{fig:dCldk}, the oscillatory behavior versus $k$ that marks the proper time distance to the Hubble crossing is largely preserved in the nearly massless case and with the largest contribution starting at high $k$ for high $l$, in general.  The massive neutrinos have a broad, peaked contribution from $k$-modes depending on the mass, with higher masses probing higher $k$ modes, and with higher $l$ modes probing higher $k$.  Fig.~\ref{fig:dCldk} shows that the $C_l$ contributions from neutrino masses of $m_\nu=0.05$~eV and less are within the linear perturbation range of $k$-modes for $l=1,2$.  The localized sensitivity to relatively high $k$ modes for low $l$ in the \CNB anisotropies for massive neutrinos may provide additional handles for constraining the scalar spectral index $n_s$ when combined with the broader $k$ mode sensitivity of the \edit{ CMB~\cite{zhang2020predictions}}.

\begin{figure}[H]
    \centering
    \includegraphics[width=0.65\linewidth]{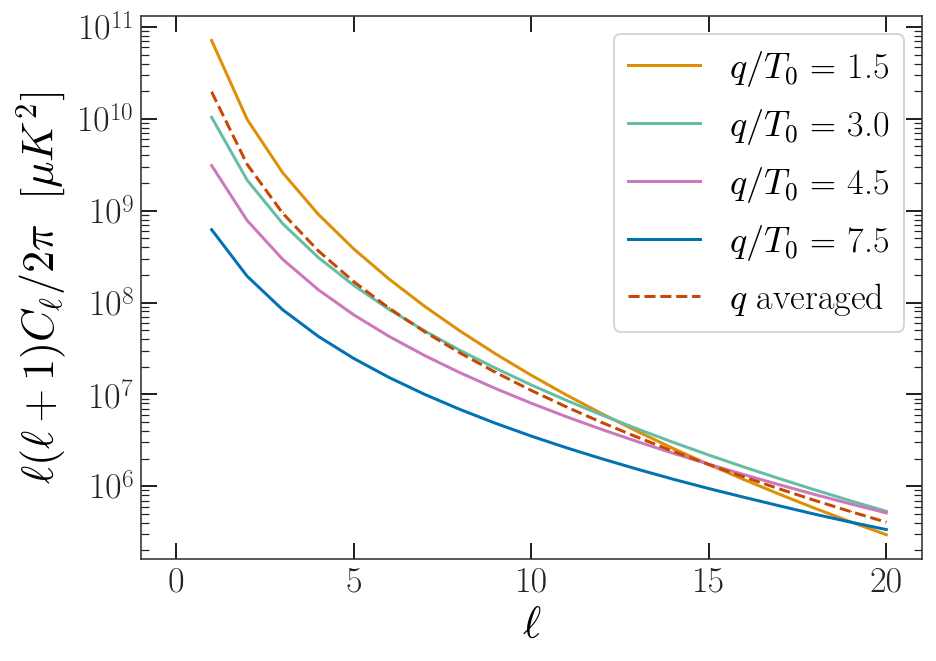}
    \caption{The $q$-averaged angular power spectra using Eq.~\eqref{eq:cl} compared with the $C_l(q)$ calculated at various $q$ values for $m_{\nu} = 0.05$~eV. The power spectra are obtained from integrating over a $k$ range of $(10^{-4}, 10^{-1})$~Mpc$^{-1}$ in logarithmic bins and using the line-of-sight integral method.}
    \label{fig:cl_qindep}
\end{figure}

The $q$-averaged angular power spectra from Eq.~\eqref{eq:cl} is more typical of the expected experimental observations for neutrino capture experiments where individual neutrino momenta are not measured directly for the range of expected relic neutrino momenta.  Though, as described in Sec.~\ref{sec:exp}, some sensitivity to the neutrino $q$-values may be present in the data.  Fig.~\ref{fig:cl_qindep} compares the $q$-averaged $C_l$ with various $q$ values for $m_\nu=0.05$~eV.  The value $q=3 T_0$ produces a similar total angular power for $l=1$ as the $q$-averaged $C_l$.  Fig.~\ref{fig:cl_qindep} shows that for a neutrino mass of 0.05~eV, the magnitude of $C_l$ drops off rapidly for increasing $q$.  The fraction of neutrinos in the Fermi-Dirac distribution that are slow moving, on order 1\% of the speed of light in the present, provide the largest contribution to the $q$-averaged $C_l$.  For a known neutrino msss, the $q$-averaged $C_l$ is therefore sensitive to the \CNB temperature $T_0$.

\begin{figure}[H]
    \centering
    \includegraphics[width=0.7\linewidth]{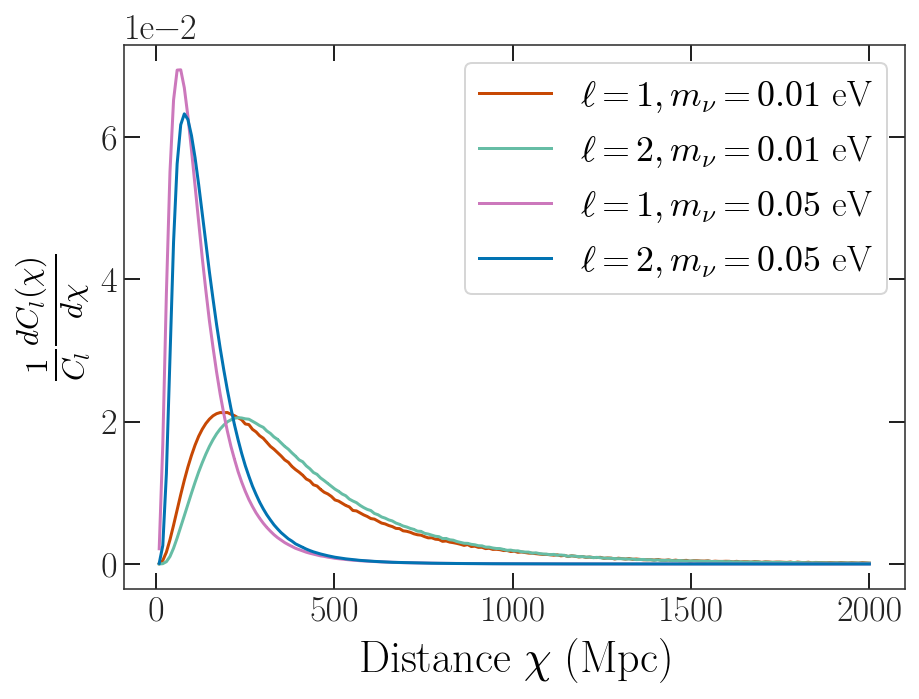}
    \caption{The normalized differential angular power spectrum as a function of the comoving distance traveled by neutrinos as defined in Eq.\eqref{eq:chi_def} for two multipole moments $l=1$ and $l=2$ and two masses. The integral over $k$ is performed over a $k$ range of $(10^{-4}, 10^{-1})$~Mpc$^{-1}$ in logarithmic bins. The $C_l$ shown in the figure are the $C_l(q)$ for $q = 3 T_0$.} 
    \label{fig:distance_pdf}
\end{figure}

The line-of-sight integral method is useful because it allows us to gain insights into the build up of \CNB anisotropies for massive neutrinos at different times and directions on the sky more conveniently than the Boltzmann hierarchy method. One of the applications of the line-of-sight method is that it allows us to calculate the fractional contribution to the total \CNB anisotropies at various times between neutrino decoupling and the present. In Fig.~\ref{fig:distance_pdf}, we show the derivatives of the angular power spectrum with respect to the distance traveled by neutrinos $\frac{d C_l}{d\chi}$ plotted at various distances.  Fig.~\ref{fig:distance_pdf} shows that similar distances are probed by $l=1,2$, while for neutrino masses $m_\nu=0.01$ and 0.05~eV, different length scales contribute to $C_l$.  The peak contributions are at low redshift.  

\section{Experimental Detection}
\label{sec:exp}

The possibility for the detection of \CNB neutrinos by the PTOLEMY experiment~\cite{betti2019design,ptolemy2019} with a polarized tritium target was first discussed in Ref.~\cite{lisanti2014measuring}. The polarization of the nuclear spin introduces an angular dependence of the neutrino capture rate with respect to the polarization direction and this, in turn, can be used to map the neutrino flux on the sky.  Angular correlations for relic neutrino capture on a more general class of nuclei are found in Ref.~\cite{akhmedov2019relic}.  The goal of the PTOLEMY detector is to count \CNB neutrinos separated by mass at the scale of 50~meV (atmospheric oscillation mass-splitting).  Since the neutrino capture rate for a given mass eigenstate depends on the fraction of electron-flavor, the maps of the neutrino sky separated by mass have intensities that depend on the neutrino mass hierarchy.

The squared amplitude for the capture of \CNB neutrinos on polarized tritium nuclei when the spin states of the electron and daughter $^3$He are not measured is proportional to
\begin{align}
\label{eq:polxsec}
    | {\cal{M}} |^2 \propto~&
    1 -   \vec{\beta}_j \cdot \vec{s}_j 
    + A (1-\vec{\beta}_j \cdot \vec{s}_j) \vec{\beta}_e \cdot \vec{s}_N \nonumber \\
    & + B K_j \vec{\beta_j} \cdot \vec{s}_N
    - B \frac{m_j}{E_j} \vec{s}_j \cdot \vec{s}_N
    + a K_j \vec{\beta}_e \cdot \vec{\beta}_j
    - a \frac{m_j}{E_j} \vec{\beta}_e \cdot \vec{s}_j
\end{align}
with
\begin{equation}
   K_j = 1 - \frac{E_j}{E_j + m_j} \vec{\beta}_j \cdot \vec{s}_j
\end{equation}
for the $j=1,2,3$ neutrino mass eigenstates, $m_j$, with $\vec{\beta}_j$ and $\vec{\beta}_e$ the neutrino $j$ and electron velocities, respectively, normalized to the speed of light and relative to the lab frame with the tritium at rest, and $\vec{s}_j$ and $\vec{s}_N$ the unit vectors for the neutrino $j$ and tritium nuclei spin directions, respectively, following the notation of~\cite{akhmedov2019relic}.  The terms in Eq.~(\ref{eq:polxsec}) proportional to the coefficients $A$ and $a$ are percent-level, while $B \simeq 1$.  For neutrino masses of order 50~meV, we can replace $\vec{\beta}_j \cdot \vec{s}_j \simeq - \beta_j$ for Dirac neutrinos and keep terms in leading order in $\beta_j$.  The modulation of the peculiar motion of the Earth enters the $\vec{\beta_j} \cdot \vec{s}_N$ term, but has little effect on the $\vec{\beta}_j \cdot \vec{s}_j$ term~\cite{lisanti2014measuring,akhmedov2019relic}.  Here we ignore the peculiar motion of the Earth.
The differential cross section for non-relativistic relic neutrino capture on polarized tritium is approximately
\begin{equation}
\label{eq:angdep}
   \frac{d\sigma}{d\Omega_j} \beta_j \simeq \frac{\bar{\sigma}_j}{4 \pi} (1 + \beta_j) (1 + \cos \alpha)
\end{equation}
with
\begin{equation}
   \bar{\sigma}_j = |U_{ej}|^2 \times 3.83 \times 10^{-45}~{\rm cm}^2
\end{equation}
where $\alpha$ is the angle between the polarization direction of the tritium nuclear spins and the velocity vector of the incoming neutrino with mass $m_j$ and $\bar{\sigma}_j$ as in~\cite{lisanti2014measuring}.  

\begin{figure}[H]
    \centering
    \includegraphics[width=0.49\linewidth]{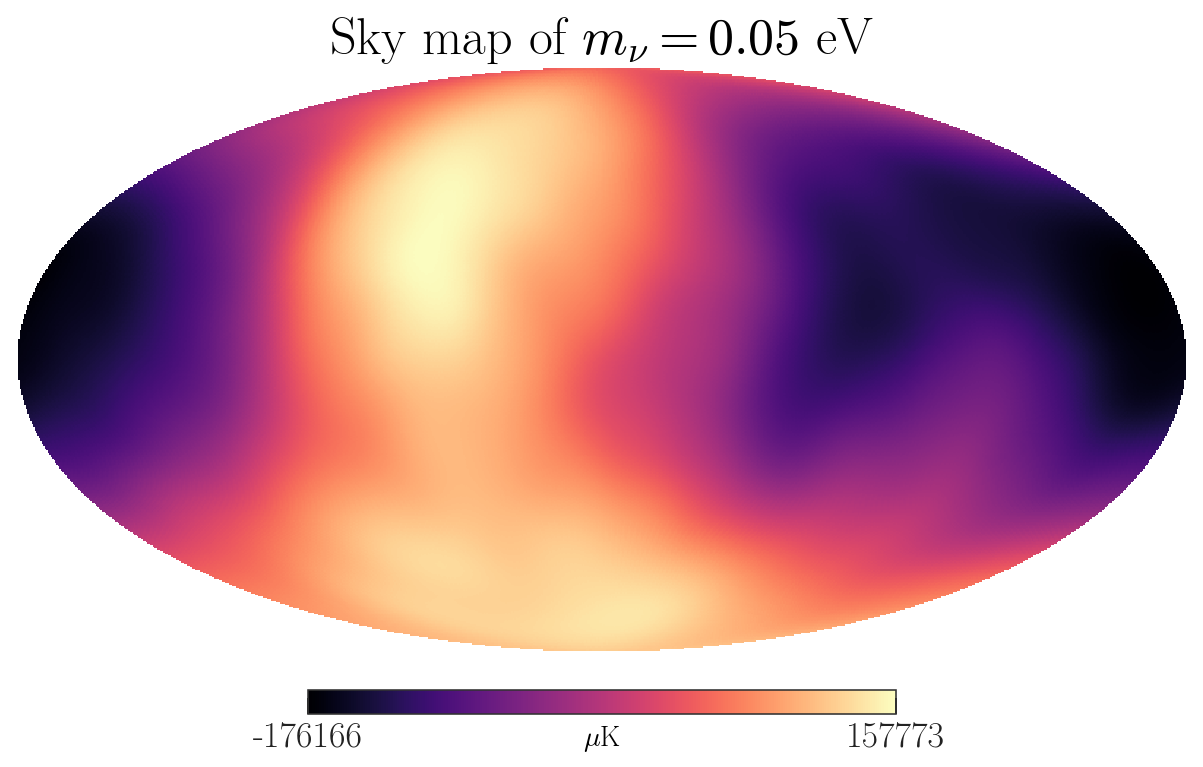}
    \includegraphics[width=0.49\linewidth]{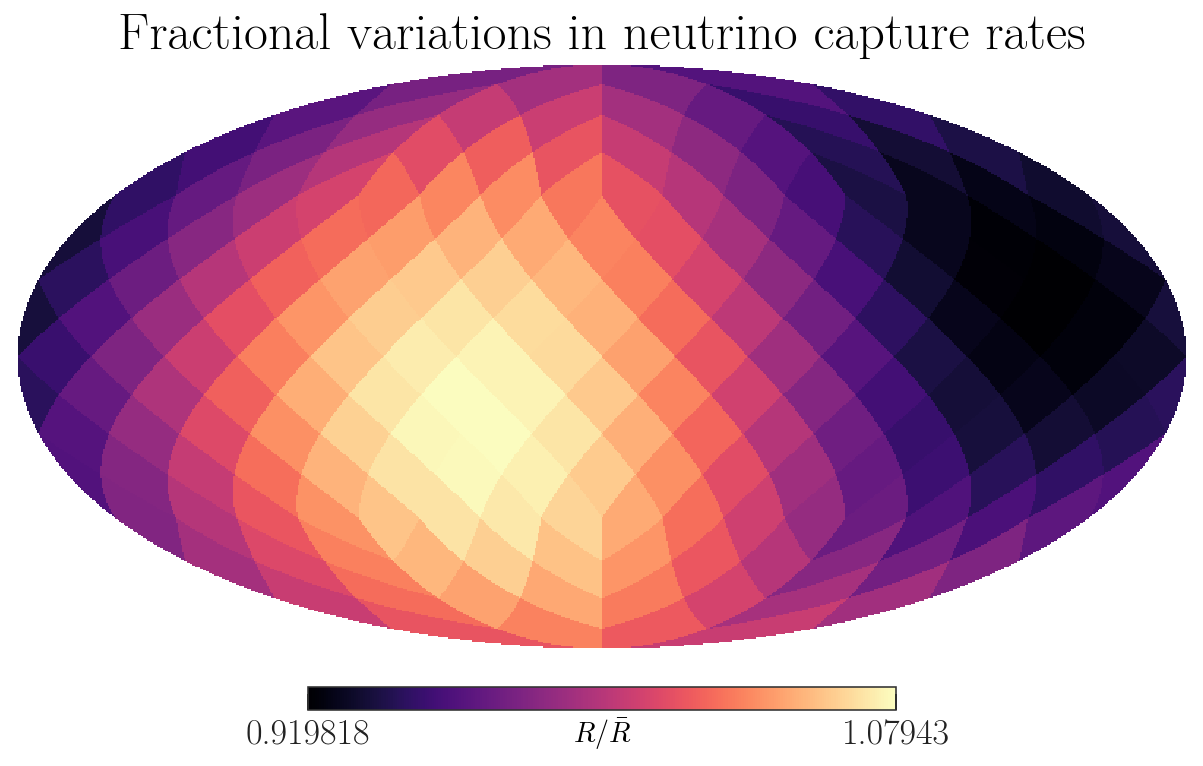}
    \caption{A sky map (\textit{left}) produced from the $q$-averaged $C_l$ with $m_\nu = 0.05$\,eV in Fig.~\ref{fig:cl_qindep} and a map of the fractional variations in the neutrino capture rate (\textit{right}) at 192 target polarization directions computed using Eq.~\eqref{eq:R}. Both maps are produced using the healpy \texttt{Python} package~\cite{Zonca2019,2005ApJ...622..759G}. }
    \label{fig:skymaps}
\end{figure}

The \CNB count rate for non-relativistic neutrino capture depends on the flux of relic neutrinos and the Dirac or Majorana nature of the neutrino with a nominal increase by a factor of 2 in count rate for Majorana over Dirac neutrinos~\cite{long2014detecting,lisanti2014measuring,roulet2018capture,akhmedov2019relic}.  
The nominal isotropic component of the neutrino flux is set by the local abundance, with the nominal prediction given in Eq.~\eqref{eq:nnu}.  The local overabundance from galactic clustering is expected to be small for neutrinos of 0.05~eV and less~\cite{ringwald2004gravitational,de2017calculation}.
 Assuming, for the moment, perfect angular resolution, the angular distribution of the neutrino flux on the sky for a given mass species $j$ can be translated in terms of a $(\theta, \phi)$-map of the temperature variations of the neutrino sky, $\delta T^j = T_0 \Theta^j(q, \theta, \phi)$, following Eqs.~\eqref{eq:FD} and \eqref{eq:nnu},
\begin{align}
    \frac{dn}{d \Omega_j} (\theta, \phi) =
    \int_0^\infty \frac{2}{(2\pi)^3} \frac{q^2 dq}{e^{\frac{q}{T_0}} + 1}
    \left(
    1 - \Gamma(q) \Theta^j(q, \theta, \phi)
    \right) .
    \label{eq:ndiff}
\end{align}
The detected relic neutrino capture rate for a fixed tritium polarization direction, $\vec{s}_N$, on the sky for a given non-relativistic Dirac neutrino species $j$ is the integral over all neutrino momenta and directions
\begin{align}
    R^j(\vec{s}_N) = 
    \iiint
    N \bar{\sigma}_j c \left(1 +  \frac{q}{m_j c}\right) 
    (1 + \cos \alpha)
    \frac{q^2 dq d\phi d(\cos\theta)}{(2\pi)^3 \left(e^{\frac{q}{T_0}} + 1\right)}
    \left(
    1 - \Gamma(q) \Theta^j(q, \theta, \phi)
    \right)
    \label{eq:R}
\end{align}
where $c$ is the speed of light, $N$ is the number of tritium nuclei in the PTOLEMY target and
\begin{align}
    \cos \alpha = \hat{n} \cdot \vec{s}_N = 
    \sin \theta_N \sin \theta \cos(\phi_N - \phi) + \cos \theta_N \cos \theta
    \label{eq:cosealpha}
\end{align}
for a unit vector $\hat{n}$ pointing at the sky with spherical polar angular coordinates $(\theta, \phi)$.  The integrals in Eqs.~\eqref{eq:ndiff} and \eqref{eq:R} are written with a $q$-dependence in  ${\Theta}^j(\vec{q})$ as the integrals over $q$ may pick up a residual dependence on the comoving neutrino momentum, even though the PTOLEMY detector is not expected to have the energy resolution to directly measure the \edit{ kinetic energies of non-relativistic neutrinos}.  
\edit{
In Fig.~\ref{fig:skymaps}, we compute the fractional variation in the neutrino capture rates predicted from Eq.~\eqref{eq:R} for the $q$-averaged $C_l$ plotted in Fig.~\ref{fig:cl_qindep} given by Eq.~\eqref{eq:cl}, for the target polarization pointing at an array of directions on the sky.  For a neutrino mass of 0.05\,eV, the neutrino capture rates are predicted to slowly vary across the sky by approximately $\pm 8$\%, relative to the nominal \CNB capture rate $\bar{R}$.  The neutrino sky anisotropies are due to the primordial power spectrum encountered in the highly amplified non-relativistic portion of the neutrino geodesic and are dominantly in low-$l$ but originating from relatively high $k$-modes, as shown in Fig.~\ref{fig:dCldk}.
Given that the largest amplification of \CNB anisotropies occur in the lowest $l$-values on the sky, the poor angular resolution of the cosine-dependence in Eq.~(\ref{eq:angdep}) is not a major limitation for measuring the angular neutrino flux spectrum on large angular scales.  Fine features of the neutrino sky map are lost in the count rate map without unfolding.
}

The direct translation of detection rates for relic neutrinos into temperature maps is further complicated by finite statistics, the dependence of the rate on local overabundance, Dirac versus Majorana neutrinos, deviations from the relativistic Fermi-Dirac distribution, the Earth's peculiar motion and other sources of modulation.  At the time of the first direct \CNB detection measurements, many of these outlying factors may be well understood or further resolved in the context of precision cosmology and neutrino physics.

\section{Conclusions \& Outlook}
\label{sec:conclusion}

For the massive neutrino \CNB, the non-relativistic transition for a given neutrino mass functions much in the same way as the matter-radiation equality functions for the CMB.  Before the non-relativistic transition, the \CNB anisotropies are small, on order $10$'s of $\mu K$ in angular power.  After the non-relativistic transition, the finite neutrino velocity allows the anisotropies to grow to roughly 10\% of the monopole temperature for $m_\nu = 0.05$~eV, picking up the largest $k$-modes on the neutrino sky.  However, the distances to the neutrino last scattering surface and the non-relativistic transition are typically an order of magnitude closer than the CMB last scattering surface and matter-radiation dominance transition~\cite{dodelson2009}.  Therefore, the massive neutrino \CNB probes high $k$ modes with a peaked sensitivity in a narrow range as compared to the broader sensitivity to $k$ modes in the CMB.  For neutrino masses of $m_\nu=0.05$~eV and smaller and for the lowest $l$-modes, the \CNB anisotropies are amplified by $k$-modes that are less than $k=0.1$~Mpc$^{-1}$ and are hence in the linear range of present-day inflationary perturbations that have been stretched across the sky.  The unique sensitivity of the \CNB anisotropies to different ranges of $k$ opens up the possibility to increase sensitivity in combination with the CMB for quantities such as the scalar spectral index $n_s$.

The statistically relevant comparison of the measured relic neutrino capture rates, as in Eq.~\eqref{eq:R}, with predictions of the angular power spectrum of the \CNB from Eq.~\eqref{eq:cl} assume that there are no preferred directions for the metric perturbations described in Eq.~\eqref{eq:boltzmann1}.  A novel aspect of the \CNB for massive neutrinos is that electromagnetic and gravitational wave observables can probe at distances of a few Gpc corresponding to the same regions of space that decoupled from relic neutrinos 13.7~billion years ago.  Estimates for the primordial metric perturbations and proper time derivatives, $\phi$ and $\partial_\tau \phi$, that go into the massive neutrino line-of-sight calculation of Eq.~\eqref{eq:delta0} can be derived \edit{along the non-relativistic portion of the neutrino geodesic} from CDM large-scale structure observations 
\editt{including the possibility to observe behind the Milky Way into the zone of avoidance}.  The comparison of precise predictions for the sky-map growth of \CNB anisotropies from CDM-tracer observables with PTOLEMY differential neutrino capture rate measurements opens up a new regime of multi-messenger astrophysics to probe inflationary perturbations in the late universe.
A self-consistent cosmology that correctly predicts the metric perturbation evolution from the time of neutrino decoupling to the present day will yield consistent results for the measured \CNB anisotropies as predicted from the physics of neutrino decoupling, neutrino properties, the expansion history and the tracers and assumed properties of dark matter and dark energy.

\acknowledgments

We thank Jeppe Mosgaard Dakin for his help with various aspects of using CO$N$CEPT. CGT is supported by the Simons Foundation (\#377485).  \edit{Some of the results in this paper have been derived using the healpy and HEALPix package.}

\bibliographystyle{JHEP}
\bibliography{references}

\providecommand{\href}[2]{#2}\begingroup\raggedright\begin{thebibliography}{10}

\bibitem{ptolemy2019}
M.G.~{Betti}, M.~{Biasotti}, A.~{Bosc{\'a}}, F.~{Calle}, N.~{Canci},
  G.~{Cavoto} et~al., \emph{{Neutrino physics with the PTOLEMY project: active
  neutrino properties and the light sterile case}},
  \href{https://doi.org/10.1088/1475-7516/2019/07/047}{\emph{\jcap} {\bfseries
  2019} (2019) 047} [\href{https://arxiv.org/abs/1902.05508}{{\ttfamily
  1902.05508}}].

\bibitem{dodelson2009}
S.~{Dodelson} and M.~{Vesterinen}, \emph{{Cosmic Neutrino Last Scattering
  Surface}}, \href{https://doi.org/10.1103/PhysRevLett.103.171301}{\emph{\prl}
  {\bfseries 103} (2009) 171301}
  [\href{https://arxiv.org/abs/0907.2887}{{\ttfamily 0907.2887}}].

\bibitem{hannestad2010}
S.~{Hannestad} and J.~{Brandbyge}, \emph{{The Cosmic Neutrino Background
  anisotropy {\textemdash} linear theory}},
  \href{https://doi.org/10.1088/1475-7516/2010/03/020}{\emph{\jcap} {\bfseries
  2010} (2010) 020} [\href{https://arxiv.org/abs/0910.4578}{{\ttfamily
  0910.4578}}].

\bibitem{michney2007}
R.J.~{Michney} and R.R.~{Caldwell}, \emph{{Anisotropy of the cosmic neutrino
  background}},
  \href{https://doi.org/10.1088/1475-7516/2007/01/014}{\emph{\jcap} {\bfseries
  2007} (2007) 014} [\href{https://arxiv.org/abs/astro-ph/0608303}{{\ttfamily
  astro-ph/0608303}}].

\bibitem{ma1995}
C.-P.~{Ma} and E.~{Bertschinger}, \emph{{Cosmological Perturbation Theory in
  the Synchronous and Conformal Newtonian Gauges}},
  \href{https://doi.org/10.1086/176550}{\emph{\apj} {\bfseries 455} (1995) 7}
  [\href{https://arxiv.org/abs/astro-ph/9506072}{{\ttfamily
  astro-ph/9506072}}].

\bibitem{oldengott2015}
I.M.~{Oldengott}, C.~{Rampf} and Y.Y.Y.~{Wong}, \emph{{Boltzmann hierarchy for
  interacting neutrinos I: formalism}},
  \href{https://doi.org/10.1088/1475-7516/2015/04/016}{\emph{\jcap} {\bfseries
  2015} (2015) 016} [\href{https://arxiv.org/abs/1409.1577}{{\ttfamily
  1409.1577}}].

\bibitem{class2011}
D.~{Blas}, J.~{Lesgourgues} and T.~{Tram}, \emph{{The Cosmic Linear Anisotropy
  Solving System (CLASS). Part II: Approximation schemes}},
  \href{https://doi.org/10.1088/1475-7516/2011/07/034}{\emph{\jcap} {\bfseries
  2011} (2011) 034} [\href{https://arxiv.org/abs/1104.2933}{{\ttfamily
  1104.2933}}].

\bibitem{2011ascl.soft02026L}
A.~{Lewis} and A.~{Challinor}, ``{CAMB: Code for Anisotropies in the Microwave
  Background}.'' \url{http://ascl.net/1102.026}, Feb., 2011.

\bibitem{schoneberg2018}
N.~{Sch{\"o}neberg}, M.~{Simonovi{\'c}}, J.~{Lesgourgues} and M.~{Zaldarriaga},
  \emph{{Beyond the traditional line-of-sight approach of cosmological angular
  statistics}},
  \href{https://doi.org/10.1088/1475-7516/2018/10/047}{\emph{\jcap} {\bfseries
  2018} (2018) 047} [\href{https://arxiv.org/abs/1807.09540}{{\ttfamily
  1807.09540}}].

\bibitem{2015Alipour}
E.~Alipour~Khayer, \emph{{The Cosmic Neutrino Background and Effects of
  Rayleigh Scattering on the CMB and Cosmic Structure}}, Ph.D. thesis, British
  Columbia U., 2015.
\newblock \url{http://doi.org/10.14288/1.0216007}.

\bibitem{Cocco:2007za}
A.G.~{Cocco}, G.~{Mangano} and M.~{Messina}, \emph{{Probing Low Energy Neutrino
  Backgrounds with Neutrino Capture on Beta Decaying Nuclei}},
  \href{https://doi.org/10.1088/1475-7516/2007/06/015}{\emph{\jcap} {\bfseries
  2007} (2007) 015}.

\bibitem{dakin2019}
J.~{Dakin}, J.~{Brandbyge}, S.~{Hannestad}, T.~{Haugb{\O}lle} and T.~{Tram},
  \emph{{{\ensuremath{\nu}}CONCEPT: cosmological neutrino simulations from the
  non-linear Boltzmann hierarchy}},
  \href{https://doi.org/10.1088/1475-7516/2019/02/052}{\emph{\jcap} {\bfseries
  2019} (2019) 052} [\href{https://arxiv.org/abs/1712.03944}{{\ttfamily
  1712.03944}}].

\bibitem{hu1995}
W.~{Hu}, D.~{Scott}, N.~{Sugiyama} and M.~{White}, \emph{{Effect of physical
  assumptions on the calculation of microwave background anisotropies}},
  \href{https://doi.org/10.1103/PhysRevD.52.5498}{\emph{\prd} {\bfseries 52}
  (1995) 5498} [\href{https://arxiv.org/abs/astro-ph/9505043}{{\ttfamily
  astro-ph/9505043}}].

\bibitem{zhang2020predictions}
G.~Zhang, ``{Predictions and Constraints on the Anisotropies in the Cosmic
  Neutrino Background}.''
  \url{http://arks.princeton.edu/ark:/88435/dsp017s75dg41g}, 2020.

\bibitem{betti2019design}
M.G.~{Betti}, M.~{Biasotti}, A.~{Bosc{\'a}}, F.~{Calle}, J.~{Carabe-Lopez},
  G.~{Cavoto} et~al., \emph{{A design for an electromagnetic filter for
  precision energy measurements at the tritium endpoint}},
  \href{https://doi.org/10.1016/j.ppnp.2019.02.004}{\emph{Progress in Particle
  and Nuclear Physics} {\bfseries 106} (2019) 120}
  [\href{https://arxiv.org/abs/1810.06703}{{\ttfamily 1810.06703}}].

\bibitem{lisanti2014measuring}
M.~{Lisanti}, B.R.~{Safdi} and C.G.~{Tully}, \emph{{Measuring anisotropies in
  the cosmic neutrino background}},
  \href{https://doi.org/10.1103/PhysRevD.90.073006}{\emph{\prd} {\bfseries 90}
  (2014) 073006} [\href{https://arxiv.org/abs/1407.0393}{{\ttfamily
  1407.0393}}].

\bibitem{akhmedov2019relic}
E.~{Akhmedov}, \emph{{Relic neutrino detection through angular correlations in
  inverse {\ensuremath{\beta}}-decay}},
  \href{https://doi.org/10.1088/1475-7516/2019/09/031}{\emph{\jcap} {\bfseries
  2019} (2019) 031} [\href{https://arxiv.org/abs/1905.10207}{{\ttfamily
  1905.10207}}].

\bibitem{long2014detecting}
A.J.~{Long}, C.~{Lunardini} and E.~{Sabancilar}, \emph{{Detecting
  non-relativistic cosmic neutrinos by capture on tritium: phenomenology and
  physics potential}},
  \href{https://doi.org/10.1088/1475-7516/2014/08/038}{\emph{\jcap} {\bfseries
  2014} (2014) 038} [\href{https://arxiv.org/abs/1405.7654}{{\ttfamily
  1405.7654}}].

\bibitem{roulet2018capture}
E.~{Roulet} and F.~{Vissani}, \emph{{On the capture rates of big bang neutrinos
  by nuclei within the Dirac and Majorana hypotheses}},
  \href{https://doi.org/10.1088/1475-7516/2018/10/049}{\emph{\jcap} {\bfseries
  2018} (2018) 049} [\href{https://arxiv.org/abs/1810.00505}{{\ttfamily
  1810.00505}}].

\bibitem{ringwald2004gravitational}
A.~{Ringwald} and Y.Y.Y.~{Wong}, \emph{{Gravitational clustering of relic
  neutrinos and implications for their detection}},
  \href{https://doi.org/10.1088/1475-7516/2004/12/005}{\emph{\jcap} {\bfseries
  2004} (2004) 005} [\href{https://arxiv.org/abs/hep-ph/0408241}{{\ttfamily
  hep-ph/0408241}}].

\bibitem{de2017calculation}
P.F.~{de Salas}, S.~{Gariazzo}, J.~{Lesgourgues} and S.~{Pastor},
  \emph{{Calculation of the local density of relic neutrinos}},
  \href{https://doi.org/10.1088/1475-7516/2017/09/034}{\emph{\jcap} {\bfseries
  2017} (2017) 034} [\href{https://arxiv.org/abs/1706.09850}{{\ttfamily
  1706.09850}}].

\bibitem{Zonca2019}
A.~Zonca, L.~Singer, D.~Lenz, M.~Reinecke, C.~Rosset, E.~Hivon et~al.,
  \emph{{healpy: equal area pixelization and spherical harmonics transforms for
  data on the sphere in Python}},
  \href{https://doi.org/10.21105/joss.01298}{\emph{Journal of Open Source
  Software} {\bfseries 4} (2019) 1298}.

\bibitem{2005ApJ...622..759G}
K.M.~{G{\'o}rski}, E.~{Hivon}, A.J.~{Banday}, B.D.~{Wandelt}, F.K.~{Hansen},
  M.~{Reinecke} et~al., \emph{{HEALPix: A Framework for High-Resolution
  Discretization and Fast Analysis of Data Distributed on the Sphere}},
  \href{https://doi.org/10.1086/427976}{\emph{\apj} {\bfseries 622} (2005) 759}
  [\href{https://arxiv.org/abs/arXiv:astro-ph/0409513}{{\ttfamily
  arXiv:astro-ph/0409513}}].

\end{thebibliography}\endgroup

\end{document}